\documentclass[singlespacing]{elsart}
\usepackage{epsfig}

\begin{document}

\begin{frontmatter}

\title{Pulse-shape discrimination with PbWO$_4$ crystal scintillators}

\author[INFN-Fi]{L.~Bardelli},
\author[INFN-Fi]{M.~Bini},
\author[INFN-Fi]{P.G.~Bizzeti},
\author[INR-Kiev]{F.A.~Danevich\thanksref{1}},
\author[INFN-Fi]{T.F.~Fazzini},
\author[INR-Kiev]{V.V.~Kobychev},
\author[MGU]{N.~Krutyak},
\author[INFN-Fi]{P.R.~Maurenzig},
\author[INR-Kiev]{V.M.~Mokina},
\author[INR-Kiev]{S.S.~Nagorny},
\author[Lviv]{M.~Pashkovskii},
\author[INR-Kiev]{D.V.~Poda},
\author[INR-Kiev]{V.I.~Tretyak},
\author[INR-Kiev]{S.S.~Yurchenko}

\address[INFN-Fi]{Dipartimento di Fisica, Universit$\acute a$ di Firenze and
INFN, 50019 Firenze, Italy}
\address[INR-Kiev]{Institute for Nuclear Research, MSP 03680 Kyiv, Ukraine}
\address[MGU]{Skobeltsyn Institute of Nuclear Physics, M.V.~Lomonosov Moscow State University, Vorob'evy Gory, 119992, Moscow, Russia}
\address[Lviv]{Department of Semiconductors Physics, Ivan Franko National University, UA-79005 Lviv, Ukraine}
\thanks[1]{Corresponding author. Address: Institute for Nuclear Research, Prospect Nauki 47, MSP 03680 Kyiv, Ukraine;
 tel: +380-44-525-1111; fax: +380-44-525-4463; e-mail address: danevich@kinr.kiev.ua}

\begin{abstract}
The light output, $\alpha/\beta$ ratio, and pulse shape have been
investigated at $-25^\circ$~C with PbWO$_4$ crystal scintillators
undoped, and doped by F, Eu, Mo, Gd and S. The fast
$0.01-0.06~\mu$s and middle $0.1-0.5~\mu$s components of
scintillation decay were observed for all the samples. Slow
components of scintillation signal with the decay times $1-3~\mu$s
and $13-28~\mu$s with the total intensity up to $\approx50\%$ have
been recognized for several samples doped by Molybdenum. We found
some indications of a pulse-shape discrimination between $\alpha$
particles and $\gamma$ quanta with PbWO$_4$ (Mo doped) crystal
scintillators.

\end{abstract}

\end{frontmatter}

\section{Introduction}

Lead tungstate crystals (PbWO$_4$) are discussed in \cite{Dane06}
as promising material for a high sensitivity experiment to search
for double beta decay of $^{116}$Cd with the help of cadmium
tungstate crystal scintillators (CdWO$_4$). PbWO$_4$ crystals can
be used as light-guides and high efficiency active shield for
low-background CdWO$_4$ scintillation detectors. Using of
radiopure PbWO$_4$ scintillators could allow to build a $2\beta$
experiment to search for $0\nu2\beta$ decay of $^{116}$Cd at the
level of sensitivity $T_{1/2}^{0\nu 2\beta}\sim10^{26}$ yr, which
corresponds to the Majorana neutrino mass $\sim0.1-0.05$ eV.

As it has been demonstrated (see, for instance
\cite{Dane03,CARVEL,Bell03}), a pulse-shape discrimination ability
of the scintillation detector is important to interpret and to
reject background caused by internal contamination of $^{232}$Th
and $^{238}$U daughters. However, pulse-shape discrimination
technique can be also useful to suppress background in a detector
from internal contamination in surrounding scintillator (for
instance, $\gamma$ background in the CdWO$_4$ detector from U/Th
contamination in the PbWO$_4$ surrounding crystals \cite{Dane06}).
Pulse-shape discrimination in PbWO$_4$ scintillators can be useful
to suppress background $\gamma$ events in CdWO$_4$ detector from
$^{208}$Tl decays in PbWO$_4$ by off line tagging $\alpha$ events
from the preceding $^{212}$Bi decay. Similarly the $\gamma$ events
from $^{214}$Bi $\beta$ decay can be rejected by tagging
subsequent $^{214}$Po $\alpha$ events in PbWO$_4$ scintillators.

The aim of the present study was to check a possibility of
pulse-shape discrimination with PbWO$_4$ crystal scintillators. In
addition, the influence of Fluorine doping on light yield of
PbWO$_4$ scintillators reported in \cite{Liu02} has been checked,
and response of crystals with different dopants to $\alpha$
particles has been investigated.

\section{Measurements and results}

\subsection{Samples}

PbWO$_4$ single crystals were grown using Czochralski technique in
an inert (Argon) atmosphere. Dopants (RE$_{3+}$ ions) were added
to the raw material (PbO$~+~$WO$_3$) in the form of RE$_2$O$_3$
oxides. Fluorine, molybdenum and sulfur were in the melt as a
PbF$_2$, MoO$_3$ and PbSO$_4$, correspondingly. Samples for
measurements were cut from crystal boules and polished in the form
of plates with a diameter of $10-15$ mm and a thickness of
$\approx5$ mm. The PbWO$_4$ crystals used in the present research
are listed in the Table 1. All the crystals were colorless and
transparent except one (6/03), which was slightly yellow colored.
Some of the samples have visible core-like defects. Concentration
of Mo, Eu, Gd dopant elements was checked by mass-spectrometry
with the help of a secondary ion mass spectrometer (Cameca,
IMS-4F). The results of the mass-spectrometry qualitatively
confirm the concentration of the dopant elements known from the
growth conditions.

\begin{table}[hb]
\caption{Parameters of the PbWO$_4$ samples.}
\begin{center}
\begin{tabular}{|l|l|l|}

\hline

 Sample & Dopants       & Concentration       \\
    ~   &   ~           & in the melt (ppm)   \\

  \hline

  20/01 & --            & --                        \\

  6/03  & F             & 400                       \\

  8/03  & F             & 4000                      \\

  9/03  & F, Gd         & 2000, 1000                \\

  7/03  & F, Eu         & 400, 100                  \\

  13/03 & F, Eu         & 2000, 20                  \\

  10/03 & F, Gd, Mo     & 1700, 90, 15800          \\

  11/03 & F, Gd, Mo     & 1000, 50, 30000          \\

  12/03 & F, Gd, Mo     & 870, 40, 78800           \\

  14/03 & F, Eu, Mo     & 1000, 6, 30000           \\

  15/03 & F, Eu, Mo, S  & 820, 5, 24500, 1000      \\

 \hline

\end{tabular}
\end{center}
\end{table}

\subsection{Relative light output}

The samples of PbWO$_4$ scintillators were wrapped by PTFE
reflector tape and optically connected to the PMT EMI9256KB by Dow
Corning Q2-3067 optical couplant. The detector was placed into
temperature controlled chamber, where the temperature of the
detector was stabilized and measured with an accuracy of $\pm
1^{\circ}$~C. The shaping time of the ORTEC (Model 572) amplifier
was set to 0.5 $\mu$s. The relative light output of the PbWO$_4$
detectors was measured  at $-25^\circ$~C with $^{60}$Co and
$^{137}$Cs $\gamma $ sources.

\nopagebreak
\begin{figure}[t]
\begin{center}
\mbox{\epsfig{figure=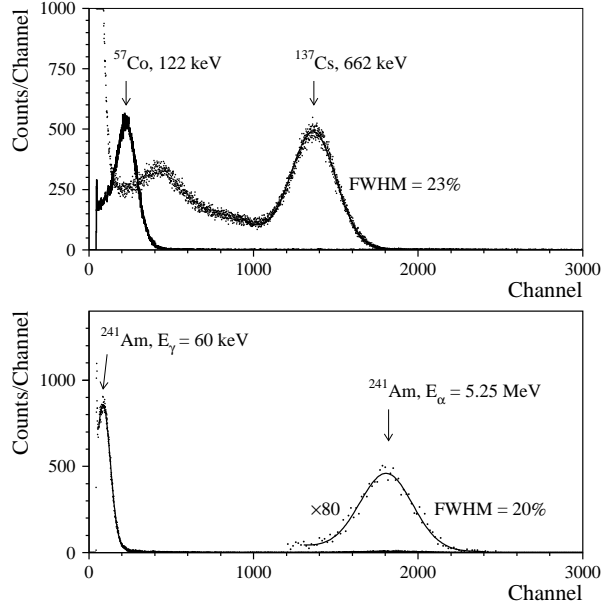,height=8.0cm}} \caption{Energy
spectra measured by the PbWO$_4$(F,Gd,Mo) scintillation crystal
(10/03) with (a) $^{57}$Co and $^{137}$Cs sources, and (b)
$^{241}$Am source at $-25^{\circ}$C.}
\end{center}
\end{figure}

Fig. 1(a) shows the energy spectra of $^{57}$Co and $^{137}$Cs
$\gamma $ sources measured with the crystal PbWO$_4$ doped by F,
Gd and Mo (10/03). An energy resolution FWHM~=~23\% was obtained
for 662 keV $\gamma $ line of $^{137}$Cs.

The spectrum measured with $^{241}$Am $\alpha$ source is shown in
Fig. 1(b). It should be stressed that even the 60 keV $\gamma$
peak of $^{241}$Am is still resolved from the PMT noise. The
energy resolution for the 60 keV $\gamma$ peak is
FWHM$~\approx~87\%$.

The samples of PbWO$_4$ scintillators were prepared from crystal
boules of different diameter ($10-15$ mm) and have values of
height-to-diameter ratio (h/D) in the range $0.3-0.5$. Relative
light output of scintillators with the high index of refraction
(the index of refraction of PbWO$_4$ crystals is 2.2) depends on
the h/D ratio. To take into account such an effect, we have
carried out measurements with CdWO$_4$ crystal scintillators
(index of refraction in the range $2.1-2.3$) with the h/D ratio in
the range $0.06-0.9$. A CdWO$_4$ crystal with the initial value of
h/D~=~0.9 was used for the measurements. To obtain the samples
with lower h/D ratio, the crystal was cut on the cleavage plane.
The CdWO$_4$ scintillators were covered by teflon reflector and
optically coupled to Philips XP2412 photomultiplier. The
measurements were carried out with a $^{207}$Bi gamma source. The
results of measurements are shown on Fig.~2. The data were fitted
by the polynomial function $f(h/D)=p_{0}+p_{1}\cdot
(h/D)+p_{2}/(h/D)$. Then the values of relative light output
measured with PbWO$_4$ scintillators were corrected by using the
experimentally determined coefficients. The relative light output
measured with the PbWO$_4$ samples after correction on the
height-to-diameter ratio (corrections do not exceed 10\%) are
presented in Table~2.

\nopagebreak
\begin{figure}[t]
\begin{center}
\mbox{\epsfig{figure=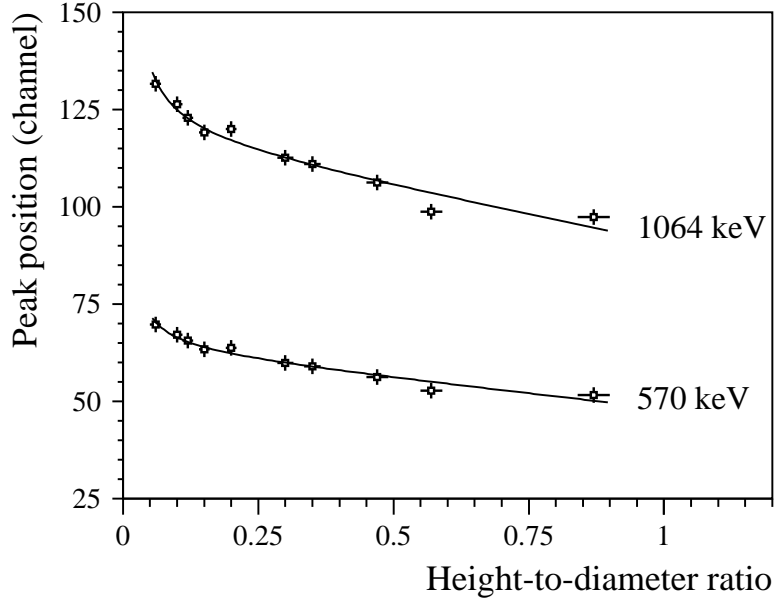,height=8.0cm}} \caption{Dependence
of relative pulse amplitude on sample height-to-diameter ratio
measured with CdWO$_4$ crystal scintillator. The solid line
represents the fits of the data by polynomial function.}
\end{center}
\end{figure}

We do not observe an increasing of light output of PbWO$_4$
scintillators grown by Czochralski method and doped by Fluorine,
as reported in \cite{Liu02}. This result is in agreement with
\cite{Koba05}. Recently a factor of 2.5 increase of the light
yield was reported for F doped PbWO$_4$ crystals grown by Bridgman
method \cite{Ye06}. In opinion of authors of Ref. \cite{Ye06}
"F-ion does not enter substantially in Czochralski-grown PWO, and
any improvement has not been observed". Therefore, additional
investigations are necessary to understand how the growth methods
affect the properties of PbWO$_4$ crystals. It should be stressed
that only initial concentrations of the dopants in the raw
materials were known in all cases. A careful quantitative control
of dopant elements, as well as impurities concentrations in the
crystals are required for such investigations.

\begin{table}[hb]
\caption{Relative light output and $\alpha/\beta$ ratio measured
with an external $^{241}$Am $\alpha$ source, and by using the
$\alpha$ peak of $^{210}$Po from internal contamination of the
PbWO$_4$ samples.}
\begin{center}
\begin{tabular}{|l|c|c|c|}
\hline

  Sample, dopants  & Relative light output  &  \multicolumn{2}{|c|} {$\alpha/\beta$ ratio} \\

  \cline{3-4}

  ~   & ~   & external $\alpha$ source & $^{210}$Po \\
  \hline

  20/01 (undoped)    & 1.00  & 0.24(3)   & 0.22(3)   \\

  6/03 (F)          & 0.94  & 0.21(3)   & 0.21(3)   \\

  8/03 (F)          & 0.98  & 0.22(3)   & 0.20(3)    \\

  9/03 (F,Gd)       & 0.81  & 0.26(4)   & 0.25(4)   \\

  7/03 (F,Eu)       & 0.73  & 0.19(4)   &  --     \\

  13/03 (F,Eu)       & 0.92  & 0.22(3)   & 0.21(3)   \\

  10/03 (F,Gd,Mo)    & 2.98  & 0.32(3)   & 0.29(3)    \\

  11/03 (F,Gd,Mo)    & 1.92  & 0.23(3)   & 0.22(3)   \\

  12/03 (F,Gd,Mo)    & 1.35  & 0.21(3)   & 0.22(3)   \\

  14/03 (F,Eu,Mo)    & 1.83  & 0.23(3)   & 0.22(3)    \\

  15/03 (F,Eu,Mo,S)  & 1.73  & 0.21(3)    & 0.22(3)   \\

 \hline

 \end{tabular}
 \end{center}
 \end{table}

\subsection{$\alpha/\beta$ ratio}

The $\alpha/\beta$ ratio was measured with the help of the
collimated $\alpha$ particles of a $^{241}$Am source. As it was
checked by surface-barrier silicon detector, the energy of
$\alpha$ particles was reduced to about 5.25 MeV going through
2~mm of air in the collimator. The energy scale was calibrated
with the help of the 570 and 1064 keV $\gamma$ lines of a
$^{207}$Bi source. The results are presented in Table ~2. The
$\alpha/\beta$ ratio measured with the external $\alpha$ source
are in the range of $0.20-0.32$ which is an agreement with result
reported in \cite{Dane06}. The errors of the $\alpha/\beta$ ratio
are mainly due to uncertainties in the energy calibration of the
PbWO$_4$ detectors due to the poor energy resolution of the
$\gamma$ peaks.

In addition, the $\alpha$ peak of $^{210}$Po from internal
contamination of crystals by $^{210}$Pb was used to measure the
$\alpha/\beta$ ratio. The energy spectrum of PbWO$_4$ scintillator
(sample 10/03) measured during 2600 s is presented in Fig.~3. A
clear peak in the spectrum can be attributed to $\alpha$ decay of
$^{210}$Po (the energy of $\alpha$ particles is 5.31 MeV) from
internal contamination of the crystal by $^{210}$Pb. The
$\alpha/\beta$ ratio is $0.29\pm 0.03$, the activity of $^{210}$Po
in the crystal is $24\pm2$ Bq/kg.

\nopagebreak
\begin{figure}[t]
\begin{center}
\mbox{\epsfig{figure=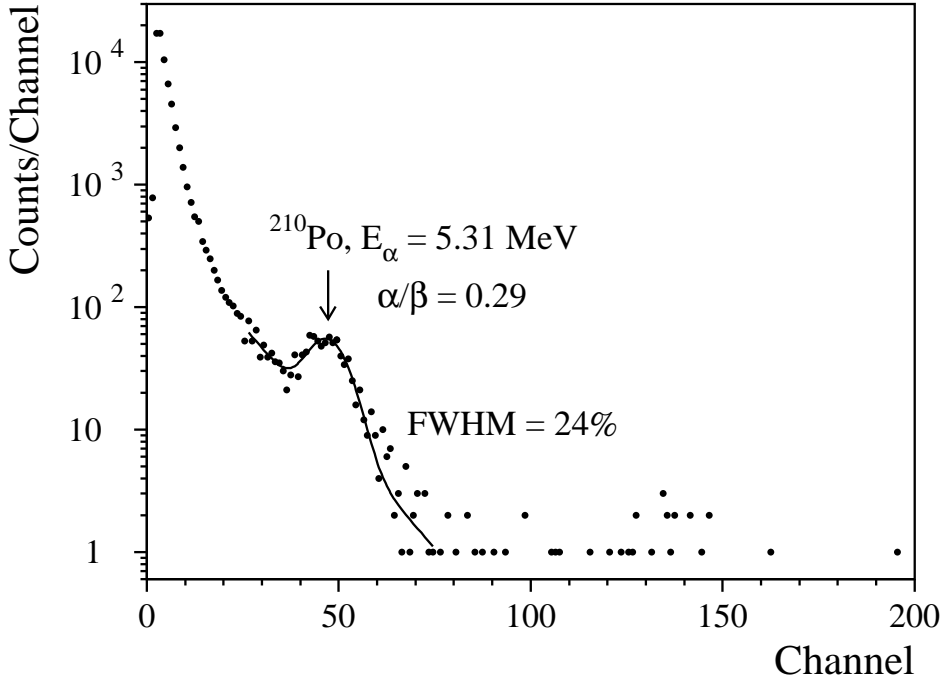,height=9.0cm}} \caption{Energy
spectrum of PbWO$_4$ scintillation crystal (10/03) measured during
2600 s. The clear peak in the spectrum at the energy  $\approx
1.54$ MeV (in $\gamma$ scale) can be attributed to $\alpha$ decay
of $^{210}$Po from internal contamination of the crystal by
$^{210}$Pb. The $\alpha/\beta$ ratio is 0.29.}
\end{center}
\end{figure}

As one can see in Table 2, the values obtained with external
$\alpha$ source and by using the peak of internal $^{210}$Po are
in agreement. Some difference in the values of the $\alpha/\beta$
ratio between the studied PbWO$_4$ samples was observed. However,
we can not interpret the difference as own properties of the
samples. It is because the $\alpha/\beta$ ratio is not only a
property of a crystal, but also a certain characteristics of the
scintillation detector depending on the shape, sizes, surface
treatment, transparency of a crystal, etc. To check a possible
effect of dopant elements on the $\alpha/\beta$ ratio in PbWO$_4$
scintillators, one has to provide the same shape, sizes,
transparency (without visible defects) and surface quality of
samples.

\subsection{Pulse-shape discrimination for $\gamma$ quanta and $\alpha$
particles}

\subsubsection{Shapes of scintillation light pulses}

The shapes of scintillation light pulses in the PbWO$_4$ crystals
were studied for $\alpha$ particles from $^{241}$Am source and for
$\gamma$ quanta from $^{137}$Cs and $^{60}$Co sources with the
help of the 125 MSample/s 12 bit digitizer described in
\cite{Pasq06}. The shapes of scintillation pulses in the sample
13/03, and in the samples doped by Molybdenum were also measured
in a time interval $\approx50~\mu$s with the help of the 20
MSample/s 12 bit digitizer \cite{Fazz98}. To determine shapes of
scintillation pulses for the sample 13/03, and for the samples
with Molybdenum, the data obtained with both the 20 and 125
MSample/s digitizers were used. The pulse shapes were constructed
in the time interval $0-1.5$ $\mu$s from the 125 MSample/s data,
and in the interval $1.5-50$ $\mu$s from the data measured by the
20 MSample/s device. The pulse shapes for $\gamma$ quanta and
$\alpha$ particles measured with the PbWO$_4$ crystal (7/03) doped
by F and Eu, and (14/03) doped by F, Eu, Mo are depicted in
Fig.~4. The pulse shape can be fitted by a sum of exponential
functions:
\begin{center}
$f(t)=\sum A_{i}(e^{-t/\tau _{i}}-e^{-t/\tau
_{0}})/(\tau_{i}-\tau_{0}),\qquad t>0$,
\end{center}
where $A_{i}$ are intensities, and $\tau_{i}$ are decay constants
for different light emission components, $\tau_{0}$ is the
integration constant of electronics ($\approx 0.02~\mu$s). The
values of $A_{i}$ and $\tau_{i}$ obtained by fitting the average
of a few thousand individual $\alpha$ and $\gamma$ pulses are
presented in Table 3.

\nopagebreak
\begin{figure}[t]
\begin{center}
\mbox{\epsfig{figure=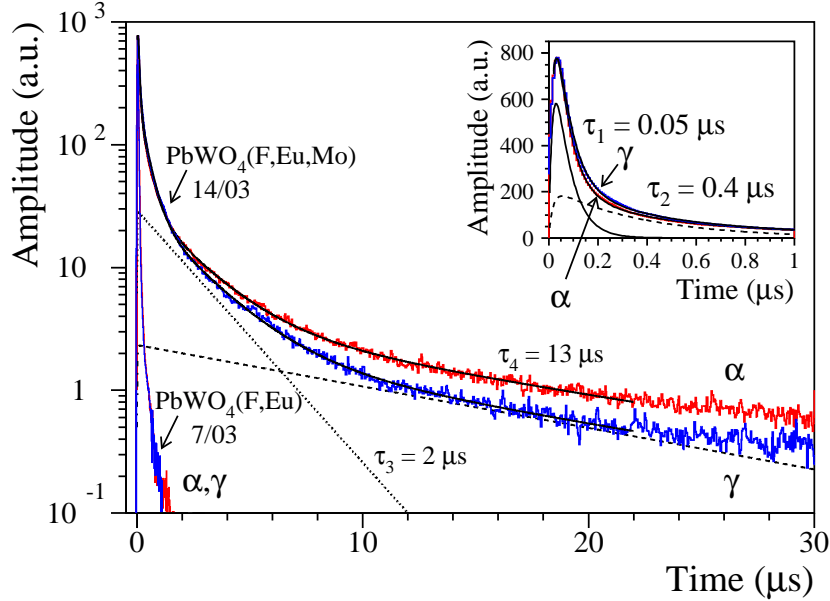,height=8.0cm}} \caption{Decay
of scintillation of the PbWO$_4$ crystal (14/03, doped by F, Eu,
Mo) for $\gamma$ quanta and $\alpha$ particles and their fit by
four components. Pulse shapes of PbWO$_4$ scintillator doped by F
and Eu (7/03) are shown for comparison. (Inset) The first part of
the PbWO$_4$ (14/03) $\alpha$ and $\gamma$ pulses.}
\end{center}
\end{figure}

\begin{table}[tbp]
\caption{The time properties of PbWO$_4$ crystal scintillators.
The decay constants and their intensities (in percentage of the
total intensity) for $\gamma$ quanta and $\alpha$ particles are
denoted as $\tau_i$ and A$_i$, respectively. The factor of merit
for pulse-shape discrimination ($FOM$, see text) characterizes the
efficiency of pulse-shape discrimination with the help of the
optimal filter (OF), and mean time (MT) methods.}
\begin{center}
\begin{tabular}{|c|c|l|l|l|l|c|c|c|}
\hline
 Sample,    & Type of       & \multicolumn{4}{|c|}{Decay constants ($\mu$s)} & \multicolumn{2}{|c|}{$FOM$} \\

 dopants    &  irradiation  & \multicolumn{4}{|c|}{and relative intensities} & \multicolumn{2}{|c|}{} \\

 \cline{3-8}

  ~         & ~             & $\tau_1$ (A$_1$)  & $\tau_2$ (A$_2$) & $\tau_3$ (A$_3$)  & $\tau_4$ (A$_4$)  & OF & MT  \\

\hline
 20/01      & $\gamma$ & 0.016 (72\%) & 0.24 (17\%) & 2.6 (11\%)   & --    & 0.05    & 0.02    \\
 (undoped)  & $\alpha$ & 0.016 (74\%) & 0.22 (16\%) & 1.9 (11\%)   & --    & ~         & ~     \\
\hline
 6/03       & $\gamma$ & 0.012 (77\%) & 0.29 (12\%) & 1.1 (11\%)   & --    & 0.08    & 0.06   \\
 (F)        & $\alpha$ & 0.013 (78\%) & 0.24 (11\%) & 1.0 (11\%)   & --    &  ~      & ~     \\
\hline
 9/03       & $\gamma$ & 0.026 (86\%) & 0.13 (14\%) & 1--5 ($<0.5\%$) & -- & 0.03   &  0.01   \\
 (F,Gd)     & $\alpha$ & 0.031 (84\%) & 0.15 (16\%) & 1--5 ($<0.5\%$) & -- & ~      & ~     \\
\hline
 7/03       & $\gamma$ & 0.014 (95\%) & 0.24 (5\%)  & 1--5 ($<0.5\%$) & -- & 0.04   & 0.01    \\
 (F,Eu)     & $\alpha$ & 0.016 (94\%) & 0.19 (6\%)  & 1--5 ($<0.5\%$) & -- &  ~     & ~     \\
\hline
 13/03      & $\gamma$ & 0.026 (79\%) & 0.5 (13\%)  & 2.0 (7\%)      & $\approx17$ (1\%) & 0.48   & 0.35 \\
 (F,Eu)     & $\alpha$ & 0.028 (84\%) & 0.4 (10\%)  & 1.8 (5\%)      & $\approx18$ (1\%) & ~      & ~     \\
\hline
 10/03      & $\gamma$ & 0.064 (42\%) & 0.30 (46\%) & 2.3 (8\%)      & 20 (4\%)     &  2.4      & 1.6  \\
 (F,Gd,Mo)  & $\alpha$ & 0.061 (33\%) & 0.37 (40\%) & 2.6 (14\%)     & 28 (12\%)    & ~         & ~     \\
\hline
 11/03      & $\gamma$ & 0.059 (25\%) & 0.40 (32\%) & 2.5 (24\%)     & 16 (19\%)  &  0.96      & 0.96  \\
 (F,Gd,Mo)  & $\alpha$ & 0.047 (23\%) & 0.36 (26\%) & 2.1 (25\%)     & 15 (26\%)    & ~         & ~     \\
\hline
 12/03      & $\gamma$ & 0.058 (25\%) & 0.38 (27\%) & 2.4 (24\%)     & 17 (24\%)    & 1.5      & 1.4   \\
 (F,Gd,Mo)  & $\alpha$ & 0.048 (22\%) & 0.35 (23\%) & 2.1 (23\%)     & 16 (32\%)    & ~         & ~     \\
\hline
 14/03      & $\gamma$ & 0.062 (25\%) & 0.38 (35\%) & 2.1 (27\%)     & 13 (13\%)   & 2.3       &  1.5  \\
 (F,Eu,Mo)  & $\alpha$ & 0.054 (24\%) & 0.35 (28\%) & 2.0 (27\%)     & 13 (21\%)    & ~         & ~  \\
\hline
 15/03       & $\gamma$ & 0.064 (26\%) & 0.38 (34\%) & 2.1 (27\%)    & 14 (12\%)      & 1.6      & 1.4  \\
 (F,Eu,Mo,S) & $\alpha$ & 0.054 (25\%) & 0.34 (28\%) & 1.7 (27\%)    & 13 (20\%)      & ~         & ~     \\
\hline

\end{tabular}
\end{center}
\end{table}

The $0.01-0.06~\mu$s and $0.1-0.5~\mu$s decay components are in
agreement with the results obtained in
\cite{Groe80,Bels95,Zhu96,Koba97,Anne98,Han00,Koba01,Zore04}. The
slow $1-3~\mu$s decay of PbWO$_4$ scintillation was also observed
in \cite{Bels95,Zhu96,Mill98}, while $13-28~\mu$s decay components
were never reported for PbWO$_4$ scintillators. We were not able
to measure the nanosecond ($\approx 1-9$ ns) decay component,
because of a comparatively large value of integration constant of
the electronics $\approx0.02~\mu$s used in the present study.

\subsubsection{Pulse-shape discrimination by optimal filter method}

A small difference in light pulse shapes could allow to
discriminate $\gamma $($\beta$) events from $\alpha$ particles. We
applied for this purpose two approaches: the optimal filter
technique proposed in \cite{Gatti} and developed in \cite{Fazz98}
for CdWO$_4$ crystal scintillator, and the mean time method.

To obtain a numerical parameter for the PbWO$_4$ signal, the
so-called shape indicator ($SI$), the following formula was
applied for each pulse:

\begin{center}
$SI=\sum f(t_k) P(t_k)/\sum f(t_k)$,
\end{center}
where the sum is over time channels $k,$ starting from the origin
of the pulse up to certain time, $f(t_k)$ is the digitized
amplitude (at the time $t_k$) of a given signal. The weight
function $P(t)$ is defined as: $P(t)=\{{f}_\alpha (t)-{f}_\gamma
(t)\}/\{f_\alpha (t)+f_\gamma (t)\}$, where $f_\alpha (t)$ and
$f_\gamma (t)$ are the reference pulse shapes for $\alpha$
particles and $\gamma$ quanta. Reasonable discrimination between
$\alpha$ particles and $\gamma$ quanta was achieved with
Molybdenum doped PbWO$_4$ crystals as one can see in Fig.~5 where
the shape indicator distributions measured by the
PbWO$_4$(F,Gd,Mo) scintillation crystal (10/03) with $\alpha$
particles ($\approx5$ MeV) and $\gamma$ quanta ($\approx~1.3$ MeV)
are shown. The small tail in the shape indicator distribution for
$\gamma$ at $\approx 50$ can be explained by background $\alpha$
events from $^{210}$Po inside the crystal (see subsection 2.3 and
Fig. 3).

As a measure of discrimination ability (factor of merit, $FOM$),
the following formula can be used:

\begin{center}
 $FOM=\mid SI_{\alpha}-SI_{\gamma}\mid/\sqrt{\sigma_{\alpha}^2+\sigma_{\gamma}^2}$,
\end{center}
where $SI_{\alpha}$ and $SI_{\gamma}$ are mean $SI$ values for
$\alpha$ particles and $\gamma$ quanta distributions (which are
well described by Gaussian functions), $\sigma_{\alpha}$ and
$\sigma_{\gamma}$ the corresponding standard deviations. For the
distributions presented in Fig.~5, the factor of merit is
$FOM=2.4$.

\nopagebreak
\begin{figure}[t]
\begin{center}
\mbox{\epsfig{figure=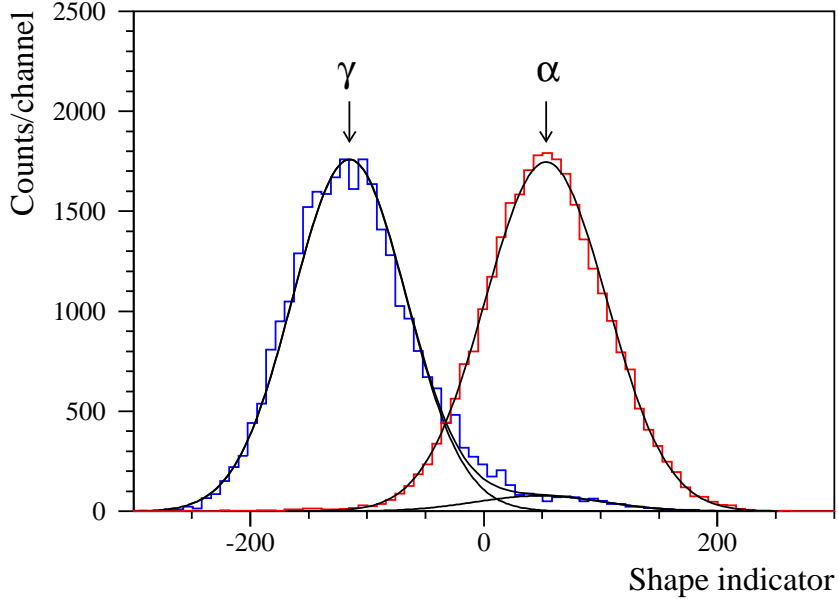,height=8.0cm}}
\caption{Distributions of the shape indicator (see text) for
pulses produced by $\gamma$ quanta and $\alpha$ particles in
PbWO$_4$(F,Gd,Mo) scintillation crystal (10/03). The small tail in
the shape indicator distribution for $\gamma$ at $\approx 50$ can
be explained by background $\alpha$ events from $^{210}$Po inside
the crystal.}
\end{center}
\end{figure}

\subsubsection{Mean time method}

The data were processed also by using the mean time method. The
following formula was applied to calculate the parameter $\langle
t\rangle$ (mean time) for each pulse:

\begin{center}
$\langle t\rangle=\sum f(t_k) t_k/\sum f(t_k)$,
\end{center}

where the sum is over time channels $k$, starting from the origin
of pulse and up to a certain time. The distributions of parameters
$\langle t\rangle$ are well described by Gaussian functions.
Therefore the factor of merit $FOM$ can be used to characterize
efficiency of pulse-shape discrimination. As one can see from
Table 3, the optimal filter method provides slightly better
pulse-shape discrimination. At the same time the mean time method
is easy to apply because it does not need the construction of a
weight function.

\section{Conclusions}

The light output, $\alpha/\beta$ ratio, and pulse shape have been
investigated at $-25^\circ$~C with PbWO$_4$ crystal scintillators
undoped, and doped by F, Eu, Mo, Gd and S. Doping of PbWO$_4$
crystals by Molybdenum improves light output by approximately a
factor of $1.5-3$. This result is an agreement with data published
before \cite{Anne00,Koba02}. The relative light output of Europium
doped PbWO$_4$ crystals is on the level of $0.7-0.9$ relatively to
undoped sample. We do not observe an increase of light output by
Fluorine doping as reported in \cite{Liu02} and \cite{Ye06}. This
fact can be explained by different methods used to grow
PbWO$_4$(F) crystals: Bridgman in \cite{Liu02,Ye06}, and
Czochralski in the present study and in the work \cite{Koba05}
where the effect was not observed too.

Values of the $\alpha/\beta$ ratio measured with different
PbWO$_4$ samples vary in the range of $0.20-0.32$. However, such a
difference can be explained by unequal quality of the samples that
have been used in the measurements. To study the possible effect
of doping on the $\alpha/\beta$ ratio one should prepare the
samples with the same sizes, surface quality, and without visible
defects.

Fast $0.01-0.06~\mu$s and middle $0.1-0.5~\mu$s components of
scintillation decay were observed for all the samples. Slow
components of scintillation signal with the decay times
$1-3~\mu$s, and $13-28~\mu$s with the total intensity at the level
of $\approx45-50\%$ have been measured for samples doped by
Molybdenum. The undoped crystal as well as crystals without
Molybdenum have shown a lower intensity of the slow components at
the level of $\leq0.5-10\%$.

We found some indications of a pulse-shape discrimination between
$\alpha$ particles and $\gamma$ quanta by applying the mean time
and optimal filter methods. The best discrimination was achieved
with PbWO$_4$ crystals doped by Molybdenum.

\end{document}